\documentclass{article}
\usepackage[T1]{fontenc} 
\usepackage[utf8]{inputenc} 
\usepackage{ismir,amsmath,cite,url}
\usepackage{amssymb}
\usepackage{graphicx}
\usepackage{color}
\usepackage{lineno}
\usepackage{multirow}
\usepackage[nolist, nohyperlinks]{acronym}

\begin{acronym}
    \acro{XAI}{explainability}
    \acro{CAV}{Concept Activation Vector}
    \acro{MIR}{Music Information Retrieval}
    \acro{NLP}{natural language processing}
\end{acronym}
\acrodefplural{CAV}[CAVs]{Concept Activation Vectors}

\title{Learning Unsupervised Hierarchies of Audio Concepts}

\multauthor
  {Darius Afchar\textsuperscript{\normalfont{1,2}} \qquad Romain Hennequin\textsuperscript{\normalfont{1}} \qquad Vincent Guigue\textsuperscript{\normalfont{2}}}
  {\textsuperscript{1}Deezer Research, France \qquad \textsuperscript{2}MLIA, Sorbonne Université, France \\ \tt{research@deezer.com}}

\def\authorname{D. Afchar, R. Hennequin, and V. Guigue}
\begin{document}

\maketitle

\begin{abstract}

Music signals are difficult to interpret from their low-level features, perhaps even more than images: e.g. highlighting part of a spectrogram or an image is often insufficient to convey high-level ideas that are genuinely relevant to humans. In computer vision, concept learning was therein proposed to adjust explanations to the right abstraction level (e.g. detect clinical concepts from radiographs). These methods have yet to be used for MIR.

In this paper, we adapt concept learning to the realm of music, with its particularities. For instance, music concepts are typically non-independent and of mixed nature (e.g. genre, instruments, mood), unlike previous work that assumed disentangled concepts.
We propose a method to learn numerous music concepts from audio and then automatically hierarchise them to expose their mutual relationships. We conduct experiments on datasets of playlists from a music streaming service, serving as a few annotated examples for diverse concepts. Evaluations show that the mined hierarchies are aligned with both ground-truth hierarchies of concepts -- when available -- and with proxy sources of concept similarity in the general case.


\end{abstract}


\section{Introduction}

Music signals are challenging to interpret \cite{liberman1968speech}. For instance, inspecting a spectrogram by parts -- \textit{e.g.} think of attention maps \cite{won2019visualizing} -- does not convey much meaning with respect to the high-level descriptions that are useful to humans -- \textit{e.g.} the mood of a song. Fortunately, a solution was proposed in a field that shares similar problems: computer vision. As pixels of images are too low-level to be understandable, \textit{concept learning} was introduced to rationalise the abstractions steps involved in a feed-forward model (\textit{e.g.} detection of colours, shapes, patterns, ...), which in turn can be used to describe the content of an image directly, and to align models with human reasonings better.  Indeed, this field has had successful applications in \ac{XAI} -- dissecting a model's predictions into human-grounded concepts \cite{kim2018TCAV, yeh2022human, ghandeharioun2022dissect}, interactive learning \cite{cai2019human} or knowledge transfer between tasks \cite{arendsen2020concept}.
However, these methods have yet to be used for \ac{MIR}.

Music has particularities that are not captured by the current formulation of concept learning.
For instance, let us consider genres as common musical concepts we would like to detect from spectrograms. Genres are typically blended as they influence or result from other genres (\textit{e.g.} \textit{Punk Rock}). By contrast, most literature on concept learning has considered sets of few identified and disentangled concepts, which does not apply to genres. Moreover, one trend in \ac{MIR} is to incorporate multiple types of descriptors into a shared space (\textit{e.g.} mood, genre, instruments \cite{hennequin2018audio, doh2020musical}), thus requiring handling even more entanglements.

We thus propose to study a novel problem for concept learning that is relevant for music: \textbf{learning hierarchies of concepts}. Indeed, a hierarchy enables navigating thousands of non-independent concepts without cognitive overload by uncovering their mutual relationships with a structure, thus rationalising which concepts share the same class or are derived from one another. In addition, the use of taxonomy -- \textit{i.e.} hierarchy -- is common in musicology and feels like a natural representation in this context \cite{von1961classification, antovic2016expectation}.

The problem we study is interesting for \textit{music representation learning} because research is often limited by data labelling, which is labour-intensive and can be ambiguous or tied to a specific dataset. Meanwhile, concepts can be learned with few shots, which open the doors to new sources of music descriptors. In particular, we propose to leverage a dataset of playlists from a music streaming service as a folksonomy of concepts. Playlists are often built with a specific concept; there exists one for any genre, instrument, mood, and many more niche ideas. Yet, there is no overall ground-truth organisation of playlists, making their use cumbersome.
That is why we propose to solve this task by first adapting a method from concept learning research to learn music concepts from playlists in a few-shot manner and then, building on top of this method, automatically derive a hierarchy from the learned concepts.

The paper proceeds as follows: we provide background on concept and hierarchy learning (section \ref{sec:prelim}); we adapt the literature of concept-based explanations to the realm of \ac{MIR} (\ref{sec:concept_learning}); we present a novel way to organise the detected concepts into a hierarchy in an unsupervised manner (\ref{sec:concept_hierarchy}); finally, we experiment and discuss our attempt to provide a new audio concepts hierarchy with mixed types (\ref{sec:exps}).


\begin{figure*}[t]
 \includegraphics[width=\textwidth]{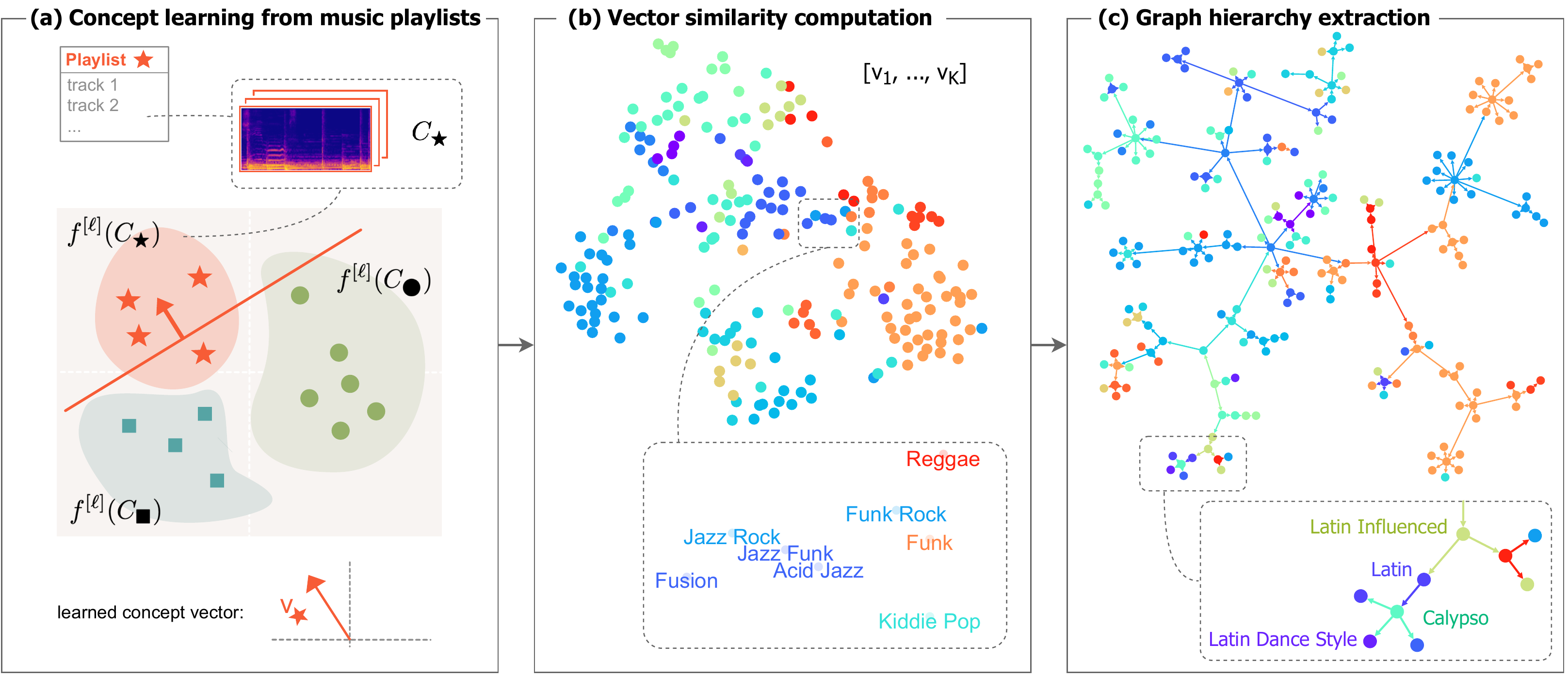}
 \caption{\textbf{Overview of our method} \textit{(a)} \acp{CAV} learning (section \ref{sec:background_cavs});  \textit{(b)} t-SNE projection of concepts $\mathcal{V}_\textrm{genre}$ learned on a small dataset with 219 music genres;  \textit{(c)} resulting extracted hierarchy $H(S(\mathcal{V}_\textrm{genre}))$. Nodes colours denote ground-truth clusters. Experiments on a bigger concept dataset are conducted in section \ref{sec:exps}.
 }
 \label{fig:overview}
\end{figure*}


\section{Preliminaries and related work}
\label{sec:prelim}

We first present the literature on concept learning and hierarchy mining, then frame our problem through \ac{XAI} to uncover common pitfalls to avoid when interpreting data.

\subsection{Concept learning}
Concept learning has witnessed growing interest in recent years \cite{yeh2022human}.
The introduction of concepts stems from the inadequacy of usual \textit{explanation spaces} with human mental models of problems: tabular features, pixels, and words in sentences are sometimes too low-level to build relevant explanations or to interact with.
This idea is far from novel from a purely technical standpoint: a stream of papers aim to learn a set of sub-attributes in a fully supervised manner (\textit{e.g.} facial features \cite{kumar2009attribute}, or animal attributes \cite{lampert2009learning}), that are then combined to predict some end-task labels, typically through a linear regression to track the importance of each attribute. This idea has been revamped recently for \textit{concept bottleneck} learning \cite{koh2020concept} and \textit{prototypical part learning} \cite{chen2019looks}. In the same vein, \textit{disentanglement techniques} \cite{zhou2018interpretable} have sought to realign latent representations to interpretable attributes. All those works rely on the assumption that a somewhat complete set of useful concepts exists, well-defined and rather disentangled, and that a corresponding labelled dataset is readily available to train on.

This is rarely the case in practice. Kim et al. introduced \acp{CAV} for images \cite{kim2018TCAV} and demonstrated that concepts could be learned reliably from few annotated examples -- technical details are provided in section \ref{sec:background_cavs}. Going beyond the previously few datasets annotated for concepts, this work has led to new and diverse applications: skin lesion interpretations \cite{lucieri2020interpretability}, emotion recognition \cite{asokan2022interpretability}, interactivity \cite{cai2019human, gopfert2022discovering}, automatic concept learning \cite{ghorbani2019towards}, batch normalisation \cite{chen2020concept}, sufficiency of explanations with shapley values \cite{strumbelj2010efficient, yeh2020completeness}, causal inference \cite{goyal2019explaining}. To our knowledge, there exists no application in the music domain.

\subsection{Unsupervised Hierarchy Mining}

Using knowledge graphs and taxonomies is frequent leverage in \ac{MIR} \cite{garcia2021leveraging}. Unsupervised learning of hierarchy is, however, less so. Part of it lies in the difficulty of adequately defining the meaning of the hierarchical relations, learning the hierarchy in a tractable way, and evaluating the found hierarchy without ground-truth.

Fortunately, there is active literature coming from \textit{\ac{NLP}} research. Many work proposed variants of \textit{topic modelling} \cite{blei2003latent} to enable sampling hierarchised structures \cite{griffiths2003hierarchical, nguyen2014learning}. However, these methods get computationally intractable as the number of nodes, topics, and hierarchical levels grows. Many successes thus came from scalable greedy methods such as hierarchical clustering \cite{lance1967general, liu2012automatic}, rule-mining \cite{volker2011statistical, anoop2016unsupervised} and graph-based analyses \cite{heymann2006collaborative, benz2010semantics}.
Outside of \ac{NLP}, hierarchies have been learned based on the latent manifold geometry \cite{ross2021benchmarks}, predefined structure \cite{jenatton2010proximal}, and for images \cite{bart2008unsupervised, hase2019interpretable}.

To our knowledge, these techniques have never been combined with concept learning, as we propose.

\subsection{Explainability}

Our end goal is to interpret music signals, which is linked to \acf{XAI}. Before jumping into our method to learn a hierarchy of concepts, we need to acknowledge that many works have debated the sanity of explanation methods \cite{doshi2017towards, adebayo2018sanity, lipton2018mythos, janzing2020feature, mahinpei2021promises} and the fact that these methods largely disagree with one another \cite{krishna2022disagreement}. XAI has vast boundaries, which has also percolated to its music sub-field for which there exists many explanation paradigms and evaluations \cite{afchar2022explainability}. Explanations for \ac{MIR} are thus also prone to exhibiting pathologies that hamper knowledge mining: \textit{e.g.} shortcut learning \cite{geirhos2020shortcut}, lack of clear definitions \cite{doshi2017towards, arrieta2020explainable}, flawed evaluations \cite{afchar2021towards}, out-of-distribution artifacts \cite{hase2021out, slack2020fooling} or misalignment with human reasonings \cite{kaur2020interpreting, dinu2020challenging, locatello2019challenging, sixt2022do}.

Uncovering these failure cases sharpens our understanding of how models make decisions, and shows that we have to go beyond evasive considerations on explainability to make them relevant \cite{lipton2018mythos} since trustworthy AI is not achieved automatically with \ac{XAI} \cite{rudin2021interpretable}.
One taxonomic distinction we would like to underline is whether methods aim to explain models or data \cite{chen2020true}. Our method is intended to belong to the latter category for which the purpose is not to explain how models work (\textit{e.g.} troubleshooting learned filters) but rather to use them as proxies to gain new insights about reality (\textit{e.g.} causal inference \cite{pearl2009causal}).

With all these aspects in mind and based on previous literature, we formalise our three explanation goals:
\begin{enumerate}
    \item \label{goal:attribution} \textbf{Attribution.} Identify musical concepts associated with a given track's audio signal.
    \item \label{goal:transferability} \textbf{Transferability.} Detected concepts should be applicable to various settings and tasks.
    \item \label{goal:generality} \textbf{Generality.} The hierarchy should be independent of the signals and make sense for various settings.
\end{enumerate}

Our first goal is linked to the traditional explanation literature supporting \textit{transparency} \cite{arrieta2020explainable, afchar2022explainability} and is common in concept-learning. We have defined our second and third goals to stress that the finality of our work is to be as universal as possible.


\section{Learning music concepts}
\label{sec:concept_learning}

In this section, we first adopt the few-shot supervision setting of concept learning to learn the music concept $\vec{v}_C$ of each playlist, viewed as a set $C$ of audio signals examples.

\subsection{Background on CAVs}
\label{sec:background_cavs}

We review some essential notions from the concept learning framework we use to learn playlist concepts \cite{kim2018TCAV}.

We denote $\vec{f}^{\,[\ell]}$ the output of the $l$-th layer of a trained neural network $f$ taking as input some samples $x \sim X$. A "concept activation vector" $\vec{v}_C$ is defined for a set of positive examples $C$ as the normal to a separating hyperplane between $\{ \vec{f}^{\,[\ell]}(x) \mid x \in C \}$ versus negative examples -- usually $\{ \vec{f}^{\,[\ell]}(x) \mid x \not \in C \}$, we provide an illustration in \figref{fig:overview}(a). In practice, the hyperplane is learned through a logistic regression on a subset of $C$, while holding out the remaining samples for validation and testing\footnote{By definition, $\vec{v}_C$ is the learned parameter of a logistic regression.}.

As additional insights, this setting is "sound" for few-shot learning because it uses a pretrained model $f$ with prior knowledge on the domain -- relating to transfer learning \cite{bozinovski1976influence} -- and learns a linear mapping on its space, hence reducing the risk of over-parametrisation and justifying why more complex mappings are not considered.

\subsection{Sets of examples $C$}

The work of Kim et al. \cite{kim2018TCAV} lends itself naturally to learning concepts from playlists. Indeed, playlists can be seen as small sets of positive music examples built around a particular musical idea. As it will be further detailed in our experiments, we can use data from music streaming services where playlists are abundant. We will denote $\mathcal{C} = \{ C_i \}_i^K$ the set of $K$ sets of music tracks that will enable to learn a \ac{CAV} set $\mathcal{V} = \{ \vec{v}_i \}_i^K$ discriminating $\mathcal{C}$ in the $\vec{f}^{\,[\ell]}$ space.

It must be noted that playlists are more than a set of tracks, we usually have access to textual data with a title and a description that help rationalise the learnt concepts.

\subsection{Backbone model $\vec{f}^{\,[\ell]}$}
\label{sec:backbone_model}

As often in few-shot learning, \acp{CAV} learning makes use of a pretrained model $f$ -- referred to as \textit{backbone} -- to embed samples into a latent space with a higher level of abstraction. 
In this paper, our working hypothesis is that \textbf{music concepts can be described solely through audio signals}. We wanted to fully explore the potential of audio signals, which is a known challenging setting in \ac{MIR} and recommendation \cite{celma2006bridging}, and leave its combination with other sources of features for future work.
In addition, having audio inputs enables using our model with any music track from any dataset, hence supporting goal \ref{goal:transferability}, while relying on features such as collaborative filtering or other usage data would have tied our model to a given dataset.

 After having experimented with several models, as the recent self-supervised model \textit{CLMR} \cite{spijkervet2021contrastive}, we have chosen to use \textit{MusiCNN} \cite{pons2019musicnn} as the backbone since it had demonstrated consistent performances across various musical tasks, which we deemed on par with our goal \ref{goal:transferability} and \ref{goal:generality}.
 MusiCNN was trained on the \textit{Million Song Dataset} \cite{bertin2011msd}.

\subsection{Preliminary results}
\label{sec:concept_discussion}

We provide some experimental observations to fix ideas and prepare the next section.
If we train \acp{CAV} as in the original paper \cite{kim2018TCAV} -- further detailed in section \ref{sec:exp_cav_learning}, we obtain a set of vectors $\mathcal{V}$ that allows to detect a set of concepts given any music input spectrogram.
An example result of learned concepts is given in \figref{fig:overview}(b) for one of the studied datasets. Quite unsurprisingly, we find that many associated concepts end up having close vector weights\footnote{In the figure, proximity is defined as the cosine similarity.}.
If we were to use our concept detectors directly, we would have many concepts activated simultaneously, which could make predictions unfathomable if $K$ is very big.
On the positive side, note that \textbf{vector proximity could be used as a means to quantify concept similarities}.
With this idea, we go a step further and next present our method to build a hierarchy to help rationalise concept dependencies.



\section{Hierarchising music concepts}
\label{sec:concept_hierarchy}

Having learned numerous non-independent concepts $\mathcal{V} = \{ \vec{v}_i \}^K_i$, we next elaborate on how to organise them with a structure to make them usable (goals \ref{goal:transferability}, \ref{goal:generality}).

Many classes of structures are possible, and it would be a mistake to select one with purely theoretical considerations since cognitive and social sciences \cite{lombrozo2016explanatory, miller2019explanation} have underlined that explanatory preferences are shaped by diverse factors that cannot be reduced to a set of so-called golden explanation principles.
From an empirical standpoint however, taxonomies seem to be frequent with music concepts (\textit{e.g.} genres \cite{hennequin2018audio}, instruments \cite{garcia2021leveraging}). We thus argue that this class of graph is rather natural for humans and chose to consider it only. 


\subsection{Similarity graph computation}

Before mining a hierarchy, it is necessary to determine how concepts -- now acting as graph nodes -- relate to one another. Essentially, we are trying to define a similarity measure between concepts. 

When sampling uniformly from a music dataset, concept activations are rare signals overshadowed by uninformative negative detection. The straightforward solution of estimating the empirical covariance of concept activations is thus not a well-behaved measure of similarity. Fortunately, there is another reliable, interpretable (and faster) way to do it. We propose to consider the positive examples of each concept playlist -- embedded in $\vec{f}^{\,[\ell]}$, and to check on what side of other concept hyperplanes their centroid lies. Formally, given $\mathcal{V} = \{\vec{v}_i\}_1^K$, we compute the matrices of concept similarities $S$ and adjacency $A$ (\textit{i.e.} $A$ is a binarised matrix representing graph edges):
\begin{align}
    \label{eq:similarity}
    S_{i,j}(\mathcal{V}) & = \mathbb{E}_{x \sim \mathcal{C}_i} \left[ \sigma ( \langle \vec{v}_j, \vec{f}^{\,[\ell]}(x) \rangle ) \right] \\
    \label{eq:adjacency}
    A_{i,j}(\mathcal{V}) & = \left[ S_{i,j}(\mathcal{V}) \geq \frac 1 2 \right]
\end{align}
where $\sigma$ denotes the logistic sigmoid function and $\mathcal{C}_i$ a random variable sampling spectrogram excerpts from tracks of the playlist $C_i$.
Bias is integrated in $\vec{v}_j$ for simplicity.
The threshold at $\frac 1 2$ for $A$ translates having each playlist centroid on the positive side of the concept hyperplanes.
Because $\mathcal{V}$ is learned through logistic regressions that return well-calibrated probabilities, $S$ and $A$ return similarities with well-defined interpretations.

\subsection{Hierarchy extraction}

Having defined similarities between concepts, we next simplify $S$ and $A$ to a hierarchy that highlights what concepts are similar to many others and what others represent unique and niche ideas.

For our problem and its scale, we propose to adapt the greedy hierarchy construction of Heymann et Garcia-Molina \cite{heymann2006collaborative} that makes use of the notion of \textbf{betweenness graph centrality} \cite{borgatti2006graph}. In this method, given an unweighted graph of similarities, a tree is greedily grown by picking each node in decreasing order of centrality in the similarity graph and linking it to the most similar and already-added node in the tree. We denote by $H(S(\mathcal{V}))$ the result of this algorithm on the similarity graph given by the adjacency matrix $A(\mathcal{V})$ and using $S(\mathcal{V})$ as similarity.

Though this algorithm was originally applied to word tags, we argue that the inductive bias of the targeted hierarchy is applicable to our problem. In particular, the use of centrality is justified by the existence of noisy links in the similarity graph, said links are assumed to be more frequent higher up in the hierarchy (\textit{general-general assumption}). For us, this means that general concepts -- \textit{e.g.} \textit{Pop} and \textit{Rock} -- should be more likely to be similar than niche concepts -- \textit{Bedroom Pop} and \textit{Goth Rock}. This sounds reasonable for music content since we expect niche concepts to relate to a specific musical idea, style, or instrument.

\begin{figure}[h]
 \center
 \includegraphics[width=\columnwidth]{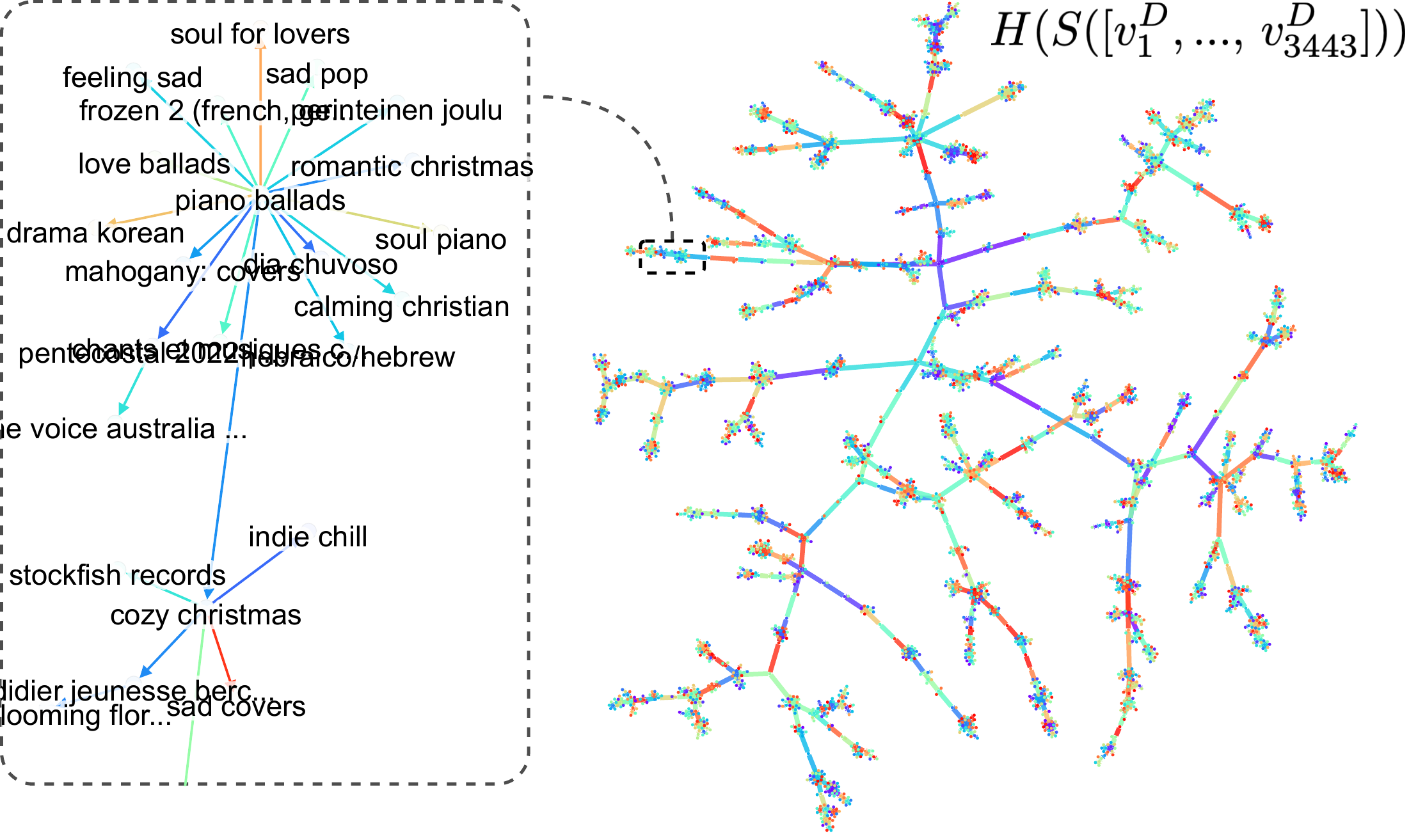}
 \caption{Obtained hierarchy on the Deezer dataset. An interactive plot of this figure, as well as many others, are available at \texttt{\color{blue} research.deezer.com/concept\_hierarchy/}.}
 \label{fig:hierarchy}
\end{figure}

\begin{table*}[h]
\begin{center}
\begin{tabular}{| c | c | c | c  c  c | }
\hline
Source & Audio ($\downarrow$) & CF ($\downarrow$) & BERT & W2V\textsubscript{1} & W2V\textsubscript{2} ($\uparrow$) \\
\hline
$H(S(\mathcal{V}_D))$
 & \textbf{2.449 $\pm$ 0.022}
 & 0.845 $\pm$ 0.013
 & 0.345 $\pm$ 0.007
 & 0.286 $\pm$ 0.009
 & 0.542 $\pm$ 0.007 \\ 
\hline
$H(S_\textrm{CF})$
 & \textbf{2.413 $\pm$ 0.021}
 & \textbf{0.345 $\pm$ 0.007}
 & 0.416 $\pm$ 0.008
 & 0.336 $\pm$ 0.010
 & 0.601 $\pm$ 0.008 \\ 
\hline
$H(S_\textrm{BERT})$
 & 2.858 $\pm$ 0.028
 & 0.868 $\pm$ 0.013
 & \textbf{0.726 $\pm$ 0.005}
 & 0.505 $\pm$ 0.011
 & 0.652 $\pm$ 0.008 \\ 
$H(S_\textrm{W2V-1})$
 & 2.952 $\pm$ 0.028
 & 0.932 $\pm$ 0.012
 & 0.523 $\pm$ 0.008
 & \textbf{0.804 $\pm$ 0.005}
 & 0.721 $\pm$ 0.007 \\ 
$H(S_\textrm{W2V-2})$
 & 2.843 $\pm$ 0.026
 & 0.847 $\pm$ 0.012
 & 0.531 $\pm$ 0.008
 & 0.596 $\pm$ 0.009
 & \textbf{0.836 $\pm$ 0.004} \\
\hline
Random
 & 3.388 $\pm$ 0.027
 & 1.104 $\pm$ 0.006
 & 0.239 $\pm$ 0.004
 & 0.142 $\pm$ 0.006
 & 0.452 $\pm$ 0.006 \\
\hline
\end{tabular}
\caption{Evaluation of the full hierarchy on audio, user-logs and title semantic similarities with 95\% confidence. \textit{Audio} and \textit{CF} source are endowed with an Euclidean similarity measure, the last three semantic sources use cosine similarities. 
}
\label{table:eval_full}
\end{center}
\end{table*}

\section{Experiments}
\label{sec:exps}

We first detail our experimental setups, then validate that concepts can be reliably learned from audio, and finally show that our proposed structuring method leads to consistent hierarchies for \ac{MIR} applications. 

\subsection{Datasets}

The set of playlists $\mathcal{C}_D$ we use was exported with the API of the streaming platform Deezer\footnote{\url{https://developers.deezer.com/api}}. We have filtered \textbf{3635 playlists of mixed types} to be used as concepts: all 1498 editorial playlists available -- considered thoughtfully curated but biased by popularity, and 2137 public user playlists with the lower artist popularity -- to improve diversity. Playlists have a mean length of 75 tracks and a minimum of 40.
This dataset includes the playlist ids, titles, descriptions, and public 30s audio preview for each track. There are 245074 unique tracks from 116497 unique artists: playlists overlap is thus very limited.

In order to compare our computed hierarchy to existing taxonomies, we make use of the APM Music dataset \cite{gao2012music}. This is a more traditionally tagged dataset, but its relevance for our task is to include a ground-truth two-level hierarchy of the \textit{music genres} (219) and \textit{moods} (165) tag types. We aggregate sets of positively tagged examples as $\mathcal{C}_\textrm{genre}$ for each genre tags and respectively $\mathcal{C}_\textrm{mood}$ for moods.

\subsection{Validating concept learning}
\label{sec:exp_cav_learning}

We first present experimental evidence related to learning sets of concepts (section \ref{sec:concept_learning}), to validate goal \ref{goal:attribution}.

\subsubsection{Experimental setup}

Concept examples are split in a 70\% (\textit{train}) / 10\% (\textit{validation}) / 20\% (\textit{test}) fashion. According to the requirements of our backbone \cite{pons2019musicnn}, we convert audio inputs to mel-spectrograms with log-magnitudes.
Representations are extracted from the penultimate layer of the backbone, denoted as $f^{[\ell]}$ -- which gave us the best performances for concept learning overall.
The complete training code with exact parameters and training regularisers is provided on our code repository for reproducibility\footnote{\color{blue} \textbf{\texttt{github.com/deezer/concept\_hierarchy}}}.




\subsubsection{Learning metrics evaluation}

\begin{figure}[h]
 \center
 \includegraphics[width=\columnwidth]{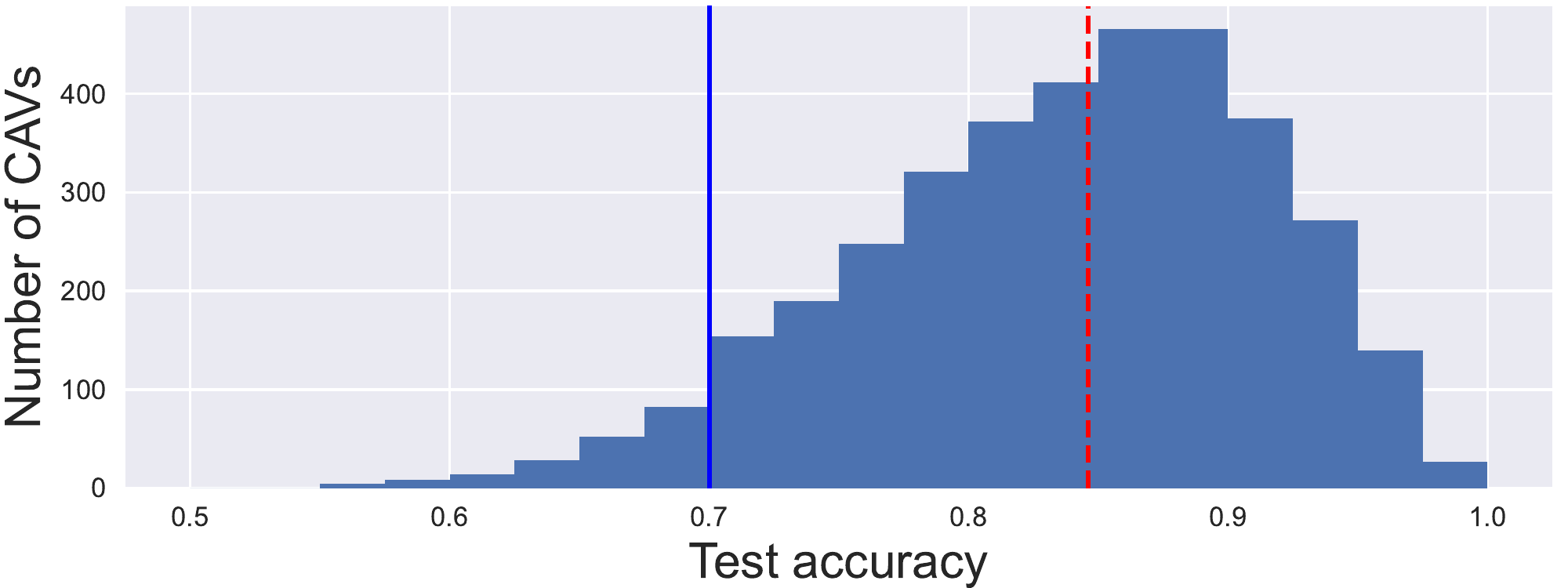}
 \caption{Concept learning performances. The left line indicates the cut-off at 70\% accuracy, the right line is the mean accuracy of the remaining \acp{CAV}.}
 \label{fig:cav_hist}
 \vspace{-0.5cm}
\end{figure}

On the Deezer dataset $\mathcal{C}_D$, the mean accuracy of learning concepts is \textbf{83.8\% $\pm$ 0.3}, with a 95\% confidence interval.
In detail, we display a histogram of test accuracies in \figref{fig:cav_hist}, showing that a \textbf{majority of playlist concepts can be learned reliably from audio}.

As a rationalisation of failure cases, the hypothesis we made in section \ref{sec:concept_discussion} to detect concepts from audio may not always hold. For instance, it fails when playlist concepts rely on factors extraneous to audio (\textit{e.g.} playlist of a movie soundtrack) or when the backbone space is not expressive enough to discriminate concepts. This actually happens in these experiments with concepts similar in all respects but for their singing languages: \textit{e.g.} the model makes no difference between Japanese and British jazz.
We filter concepts $v_i$ below a given test accuracy threshold to account for this effect. Fixed at 70\%, 192 concepts are filtered out, leading to $K=3443$ remaining concepts. This threshold was set empirically by checking that those left-out concepts were indeed unclear from audio (\textit{e.g.} OST, new releases, charts). We denote the resulting concept set $\mathcal{V}_D$.

The two smaller APM datasets lead to similar results with a \textbf{78.3\% $\pm$ 0.8} for genres, denoted $\mathcal{V}_\textrm{genre}$, and a \textbf{71.3\% $\pm$ 0.6} accuracy for mood tags, denoted $\mathcal{V}_\textrm{mood}$.

As a wrap-up of this section, goal \ref{goal:attribution} of detecting concepts from audio seems satisfied in our experiments. 

\subsection{Hierarchy Mining}

We conduct a three-fold evaluation of our proposed hierarchy extraction method (section \ref{sec:concept_hierarchy}) in order to evaluate goal \ref{goal:generality}: structure; ground-truth comparison; alignment with other sources of concept similarities.

\subsubsection{Structure priors}
\label{sec:exp_strct}

We first provide evidence supporting our choice to extract a hierarchical tree graph over other choices of structures. Though we do not have access to a ground-truth graph of the relations between concepts in $\mathcal{C}_D$, we have some priors on what a satisfying result should be like. For instance, we know that musical concepts are often blended (\textit{e.g.} genres). It is thus safe to assume that a good graph of concepts should be \textit{connected}. Extending this property, isolated nodes should strongly be discouraged. As another prior assumption on our target graph, we know that we can usually find several similar playlists to a given one, but that very strongly connected nodes should be infrequent, \textit{sparsity} should thus also be valued.

With these two priors in mind, we inspect various graph baselines that could straightforwardly be obtained from $\mathcal{V}_D$ without our hierarchy extraction step: the similarity graph $A(\mathcal{V}_D)$ from equation \eqref{eq:adjacency}; a similarity graph obtained from $S(\mathcal{V}_D)$ with an adjusted threshold to match the sparsity of $H(S(\mathcal{V}_D))$ (\textit{Sparse $A$}); a top-1 most similar neighbours graph of each node; reference random graphs.
Their respective sparsity and connectivity is shown in \tabref{table:struct}. It appears that baseline graphs become disconnected as soon as sparsity drops. We also underline that random baselines provide better-behaved graphs than their counterparts, indicating that similarity links are not evenly distributed among nodes, thus pointing to the existence of communities in the graph that justifies the use of betweenness \cite{girvan2002community}. Using a hierarchy extraction algorithm helps maintain a low sparsity while connecting every node in the graph.

\begin{table}[h]
\begin{center}
\small
\begin{tabular}{| c| c | c c |  }
\hline
Graph & \#Edges ($\downarrow$) & \#CC ($\downarrow$) & \#IN ($\downarrow$)  \\
\hline
$H$ & 3442 (0.03\%) & 1 & 0 \\ 
\hline
$A$ & 1309163 (11\%) & 1 & 0 \\ 
Sparse $A$ & 3443 & 2766 & 2735 \\ 
Top-1 & 3443 & 134 & 0 \\ 
\hline
Random Sparse $A$ & 3443 & 516 & 430 \\ 
Random Top-1 & 3443 & 7 & 0 \\ 
\hline
\end{tabular}
\caption{Structure evaluation. We count the number of edges in each graph,
of connected components (\textit{CC}), and of isolated nodes (\textit{IN}).}
\label{table:struct}
\end{center}
\vspace{-0.3cm}
\end{table}

This study does not prevent using sparse and connected graphs other than hierarchies. We believe that other relevant structures may exist. Nonetheless, as additional benefits of using hierarchies, we note that tree-like graphs are always planar, which eases their visualisation and navigability by having only one most-relevant parent per node.


\subsubsection{Ground-truth evaluation}
\label{sec:exp_apm}

To validate our generated concept hierarchy, we compare $H(S(\mathcal{V}_\textrm{mood}))$ and  $H(S(\mathcal{V}_\textrm{genre}))$ to the expert tag hierarchy provided with the APM dataset. Note that this unavailable for $\mathcal{V}_D$.
As those hierarchies are two-levelled  -- \textit{i.e.} clustering of tags, we evaluate the accuracy of the mined edges at linking neighbours from the same cluster and compute a silhouette score \cite{rousseeuw1987silhouettes}. Judging from the results given in \tabref{table:eval_apm}, the generation is promising but far from perfect.

The estimated hierarchy of genres is illustrated in \figref{fig:overview}(c) to help interpret the results. As a typical failure case, the orange cluster ("Electronica") ends up in two separate branches, which can happen because of the greedy construction of our hierarchy.
In other cases, however, the ground-truth may be questioned: "Latin influence", "Latin", "Ethnic Dance" and "Calypso" form a link in our hierarchy, but belong to four different clusters in the ground-truth -- due to non-musical taxonomic considerations -- despite being musically similar.
It is not easy to quantify the performance upper-bound we have between our musical concept detection and the practical expert taxonomies we compare to. Nevertheless, seeing that our generated hierarchies roughly align with ground-truth structures is a good sanity check.

\begin{table}[h]
\small
\begin{center}
\begin{tabular}{| c | c | c | }
\hline
Tags & Accuracy (\%) & Silhouette \\
\hline
$H_\textrm{mood}$ & 45.1 & -0.09 $\pm$ 0.05  \\
$H_\textrm{genre}$ & 49.1 & -0.17 $\pm$ 0.04   \\

\hline

\end{tabular}
\caption{Evaluation of our unsupervised hierarchy against ground-truth clustering of moods and genres tags.}
\label{table:eval_apm}
\end{center}
\vspace{-0.3cm}
\end{table}


\subsubsection{Alignment to various sources}

Coming back to the difficult setting of $\mathcal{V}_D$, we finally evaluate our hierarchy of concepts with mixed types, illustrated in \figref{fig:hierarchy}. As we do not have a ground-truth in this general case, we evaluate whether the edges of our hierarchy -- built from $S(\mathcal{V})$ -- make sense on other sources of similarity.
Specifically, since our data is collected from a streaming service, collaborative filtering embeddings \cite{hu2008collaborative} based on listening logs are available to estimate playlist similarities ($S_\textrm{CF}$). Playlists similar in concepts could indeed be expected to be co-listened by users, though popularity biases are also at play.
We also leverage playlist titles and expect neighbours to be rather close semantically. To that end, we use a large language model \cite{reimers2019sentence} that can embed any text prompt ($S_\textrm{BERT}$), and two music-specialised word embedders for which we average representations of in-vocabulary words of each concept names: $S_\textrm{W2V-1}$ \cite{doh2020musical} and $S_\textrm{W2V-2}$ \cite{won2021multimodal}. For reference, we generate hierarchies from each domain-specific source, include a random hierarchy, and a \textit{Audio} measure based on \acp{CAV} weights, following our observations made in section \ref{sec:concept_discussion}.

We compute the average similarity on each hierarchy's edges given those several sources of similarities. Results are provided in \tabref{table:eval_full}. By construction, each hierarchy maximises performance on the source it was built on. We rather inspect how one source of knowledge transfers to other sources. Collaborative filtering being the canonical way of estimating similarities in the streaming industry our dataset is extracted from and in general \cite{afchar2022explainability}, we underline that our hierarchy -- solely based on audio -- closely matches its transfer performances, which is a good result. We have thus shown that our hierarchy transfers to other sources of knowledge (goal \ref{goal:generality}).

\subsection{Qualitative discussion}

We retrieve many associations that make sense for humans in $H(S(\mathcal{V}_D))$: blues and jazz; rock and pop; motivation, dancing, running, and party are close to one another -- which aligns with previous work \cite{ibrahim2020audio}. More interestingly, we observe that \textit{"LSD Trip"} is the parent of \textit{"Surf"}, \textit{"Summer of love"}, and \textit{"Stoner Rock"}, effectively associating an activity to a sport, an historic phenomenon, and a music genre. Yet, we also find some association that make sense for the model but not for humans: \textit{e.g.} \textit{"Rock Christmas"} is also a child of \textit{"LSD Trip"}. This phenomenon is hard to avoid with fully unsupervised methods \cite{locatello2019challenging, dalvi2022discovering}.
We can find many more interesting associations, but we have to beware of confirmation biases, as often in \ac{XAI} \cite{nickerson1998confirmation, dinu2020challenging, kaur2020interpreting}. We provide online visualisations of the results\footnote{\color{blue} \textbf{\texttt{research.deezer.com/concept\_hierarchy/}}}.

As observable limitations, $\mathcal{V_D}$ is biased by the streaming platform usage towards French, English, and Brazilian content, which could be addressed in future work with importance sampling. Then, some concept titles are misleading, \textit{e.g.} playlists "City sounds: <city>" are confounded by the taste of their curator, resulting in city concepts being close to the 60's genre, despite this not being obvious solely judging from the title. Finally, as already mentioned, the backbone fails to discriminate linguistic information: \textit{e.g.} \textit{"Queer Pop"} and \textit{"[Current] Pop"} differ by their lyric theme but are neighbours in the hierarchy.

\section{Conclusion}

Spectrograms are hard to interpret. We propose to extend concept learning to learn hierarchies of music concepts from playlists, with greater flexibility than usual music taggers. Our results should be viewed as complementary to expert music ontologies, as a means to witness how music is organically described by users and editors, and thus to capture new or evolving salient aspects of music.

This topic is novel and further steps are necessary in order to to overcome cultural biases of our data, and to discover causal relations for our structure. Future work include leveraging the found hierarchy for dynamic music recommendation -- e.g. exploring and switching branches.

\clearpage

\bibliography{refs.bib}

\end{document}